\documentclass{article}
\usepackage{graphics,pstricks,pst-node}
\title{Complexity of Networks (reprise)}

\author{Russell K. Standish\\
                Mathematics and Statistics, University of New South
                Wales\\
                hpcoder@hpcoders.com.au
        }

\def\citeyear(#1)#2{\cite{#2}}

\newcommand{\EcoLab}{{\sffamily\slshape
    \mbox{\raisebox{.5ex}{Eco}\hspace{-.4em}{\makebox[.5em]{L}ab}}}}

\begin{document}
\maketitle
\begin{abstract}
Network or graph structures are ubiquitous in the study of complex
systems. Often, we are interested in complexity trends of these system
as it evolves under some dynamic. An example might be looking at the
complexity of a food web as species enter an ecosystem via migration
or speciation, and leave via extinction.

In a previous paper, a complexity measure of networks was proposed based on
the {\em complexity is information content} paradigm. To
apply this paradigm to any object, one must fix two things: a representation
language, in which strings of symbols from some alphabet describe, or
stand for the objects being considered; and a means of determining
when two such descriptions refer to the same object. With these two
things set, the information content of an object can be computed in
principle from the number of equivalent descriptions describing a
particular object.

The previously proposed representation language had the deficiency
that the fully connected and empty networks were the most complex for
a given number of nodes. A variation of this measure, called
zcomplexity, applied a compression algorithm to the resulting bitstring
representation, to solve this problem. Unfortunately, zcomplexity
proved too computationally expensive to be practical.

In this paper, I propose a new representation language that encodes
the number of links along with the number of nodes and a representation
of the linklist.  This, like zcomplexity, exhibits minimal complexity
for fully connected and empty networks, but is as tractable as the
original measure.

This measure is extended to directed and weighted links, and several
real world networks have their network complexities compared with
randomly generated model networks with matched node and link counts,
and matched link weight distributions. Compared with the random
networks, the real world networks have significantly higher
complexity, as do artificially generated food webs created via an
evolutionary process, in several well known ALife models.


\end{abstract}

\section{Introduction}

This work situates itself firmly within the {\em complexity is
information content} paradigm, a topic that dates back to the 1960s
with the work of Kolmogorov, Chaitin and Solomonoff. Prokopenko et
al. \cite{Prokopenko-etal09} provide a recent review of the area, and
argue for the importance of information theory to the notion of
complexity measure.  In \cite{Standish01a}, I argue that {\em
information content} provides an overarching complexity measure that
connects the many and various complexity measures proposed (see
\cite{Edmonds99} for a review) up to that time. In some ways,
information-based complexity measures are priveleged, motivating the
present work to develop a satisfactory information-based complexity
measure of networks.

The idea is fairly simple.  In most cases, there is an obvious {\em
prefix-free} representation language within which descriptions of the
objects of interest can be encoded.  There is also a classifier of
descriptions that can determine if two descriptions correspond to the
same object. This classifier is commonly called the {\em observer},
denoted $O(x)$.

To compute the complexity of some object $x$, count the number of
equivalent descriptions $\omega(\ell,x)$ of length $\ell$ that map to
the object $x$ under the agreed classifier. Then the complexity of $x$
is given in the limit as $\ell\rightarrow\infty$:
\begin{equation}\label{complexity}
{\cal C}(x) = \lim_{\ell\rightarrow\infty} (\ell\log N - \log\omega(\ell,x))
\end{equation}
where $N$ is the size of the alphabet used for the representation language.

Because the representation language is prefix-free, every description
$y$ in that language has a unique prefix of length $s(y)$.  The
classifier does not care what symbols appear after this unique prefix.
Hence $\omega(\ell,O(y))\geq N^{\ell-s(y)}$. As $\ell$ increases,
$\omega$ must increase as fast, if not faster than $N^\ell$, and do so
monotonically. Therefore $C(O(y))$ decreases monotonically with
$\ell$, but is bounded below by 0. So equation (\ref{complexity})
converges.

The relationship of this algorithmic complexity measure to more
familiar measures such as Kolmogorov (KCS) complexity, is given by the
coding theorem\cite[Thm 4.3.3]{Li-Vitanyi97}. In this case, the
descriptions are halting programs of some given universal Turing
machine $U$, which is also the classifier. Equation (\ref{complexity})
then corresponds to the logarithm of the {\em universal a priori
probability}, which is a kind of formalised Occam's razor that gives
higher weight to simpler (in the KCS sense) computable theories for
generating priors in Bayesian reasoning. The difference between this
version of ${\cal C}$ and KCS complexity is bounded by a constant
independent of the complexity of $x$, so these measures become
equivalent in the limit as message size goes to infinity.

Many measures of network properties have been proposed, starting with
node count and connectivity (no. of links), and passing in no         
particular order through cyclomatic number (no. of independent loops),
spanning height (or width), no. of spanning trees, distribution of
links per node and so on. Graphs tend to be classified using these
measures --- {\em small world} graphs tend to have small spanning
height relative to the number of nodes and {\em scale free} networks
exhibit a power law distribution of node link count.

Some of these measures are related to graph complexity, for example
node count and connectivity can be argued to be lower and upper bounds
of the network complexity respectively. More recent attempts include
{\em offdiagonal complexity}\cite{Claussen04}, which was compared with
an earlier version of this proposal in \cite{Standish05a}, and {\em
medium articulation}\cite{Wilhelm-Hollunder07, Kim-Wilhelm08}.

However, none of the proposed measures gives a theoretically
satisfactory complexity measure, which in any case is context
dependent (ie dependent on the {\em observer} $O$, and the
representation language).

In setting the classifier function, we assume that only the graph's
topology counts --- positions, and labels of nodes and links are not
considered important. Links may be directed or undirected. We consider
the important extension of weighted links in a subsequent section.

A plausible variant classifier problem might occur when the network
describes a dynamical system (such as interaction matrix of the EcoLab
model, which will be introduced later). In such a case, one might be
interested in classifying the network according to numbers and types
of dynamical attractors, which will be strongly influenced by the
cyclomatic count of the core network, but only weakly influenced by
the periphery. However, this problem lies beyond the scope of this
paper.

The issue of representation language, however is far more problematic.
In some cases, eg with genetic regulatory networks, there may be a
clear representation language, but for many cases there is no uniquely
identifiable language. However, the {\em invariance theorem}\cite[Thm
2.1.1]{Li-Vitanyi97} states that the difference in complexity
determined by two different {\em Turing complete} representation
languages (each of which is determined by a universal Turing machine)
is at most a constant, independent of the objects being measured.
Thus, in some sense it does not matter what representation one picks
--- one is free to pick a representation that is convenient, however
one must take care with non-Turing complete representations.

In the next section, I will present a concrete graph description
language that can be represented as binary strings, and is amenable to
analysis. The quantity $\omega$ in eq (\ref{complexity}) can be simply
computed from the size of the automorphism group, for which
computationally feasible algorithms exist\cite{McKay81}.

The notion of complexity presented in this paper naturally marries
with thermodynamic entropy $S$\cite{Layzer88}:
\begin{equation}\label{Layzer}
S_{\max}={\cal C} + S
\end{equation}
where $S_{\max}$ is called {\em potential entropy}, ie the largest
possible value that entropy can assume under the specified
conditions. 
The interest here is that a dynamical process updating
network links can be viewed as a dissipative system, with links being
made and broken corresponding to a thermodynamic flux. It would be
interesting to see if such processes behave according the maximum
entropy production principle\cite{Dewar03} or the minimum entropy
production principle\cite{Prigogine80}.

In artificial life, the issue of complexity trends in evolution is
extremely important\cite{Bedau-etal00}. I have explored the complexity
of individual Tierran organisms\cite{Standish03a,Standish04c}, which,
if anything, shows a trend to simpler organisms. However, it is
entirely plausible that complexity growth takes place in the network
of ecological interactions between individuals. For example, in the
evolution of the eukaryotic cell, mitochondria are simpler entities
than the free-living bacteria they were supposedly descended from. A
computationally feasible measure of network complexity is an important
prerequisite for further studies of evolutionary complexity trends.

In the results section, several well-known food web datasets are
analysed, and compared with randomly-shuffled ``neutral models''. An
intriguing ``complexity surplus'' is observed, which is also observed
in the food  webs of several different ALife systems that have
undergone evolution.

\section{Representation Language}

One very simple implementation language for undirected graphs is to
label the nodes $1..n$, and the links by the pair $(i,j), i<j$ of
nodes that the links connect. The linklist can be represented simply
by an $L=n(n-1)/2$ length bitstring, where the $\frac12j(j-1)+i$th
position is 1 if link $(i,j)$ is present, and 0 otherwise.

The directed case requires doubling the size of the
linklist, ie or $L=n(n-1)$. We also need to prepend the string with
the value of $N$ in order to make it prefix-free --- the simplest
approach is to interpret the number of leading 1s as the number $n$,
which adds a term $n+1$ to the measured complexity.

This proposal was analysed in \cite{Standish05a}, and has the
unsatisfactory property that the fully connected or empty networks are
maximally complex for a given node count. An alternative scheme is to
also include the link count as part of the prefix, and to use binary
coding for both the node and link counts. The sequence will start with
$\lceil\log_2n\rceil$ 1's, followed by a zero stop bit, so the prefix will be
$2\lceil\log_2n\rceil+\lceil\log_2L\rceil+1$ bits.

This scheme entails that some of bitstrings are not valid networks,
namely ones where the link count does not match the number of 1s in
the linklist. We can, however, use rank encoding\cite{Myrvold-Ruskey01} of the
linklist to represent the link pattern. The number of possible
linklists corresponding to a given node/link specification is given by
\begin{equation}
\Omega = \left(\begin{array}{c} 
L \\
l \\
\end{array}
\right) = \frac{L!}{(L-l)! l!}
\end{equation}
This will have a minimum value of 1 at $l=0$ (empty network) and
$l=L$, the fully connected network.

Finally, we need to compute $\omega$ of the linklist, which is just
the total number of possible renumberings of the nodes ($n!$), divided by
the size of the graph automorphism group, which can be practically computed by
Nauty\cite{McKay81}, or a new algorithm I developed called
SuperNOVA\cite{Standish09a} which exhibits better performance on
sparsely linked networks.

\begin{figure}
\begin{center}
\resizebox{\textwidth}{!}{\includegraphics{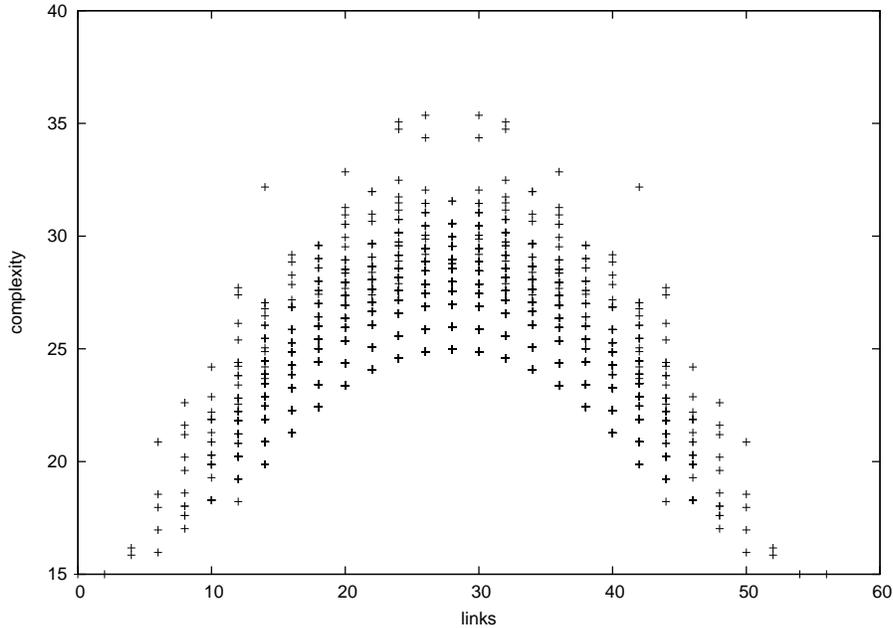}}
\end{center}
\caption{The new complexity measure as a function of link count for
  all networks with 8 nodes. This shows the strong dependence of
  complexity on link count, and the symmetry between networks and
  their complements.
}
\label{l-C}
\end{figure}

A network $A$ that has a link wherever $B$ doesn't, and vice-versa
might be called a complement of $B$. A bitstring for $A$ can be found
by inverting the 1s and 0s in the linklist part of the network
description. Obviously, $\omega(A,L)=\omega(B,L)$, so the complexity
of a network is equal to that of its complement, as can be seen in
Figure \ref{l-C}.

A connection can be drawn from this complexity measure, to that
proposed in \cite{Dorin-Korb10}. In their proposal, nodes are
labelled, so the network is uniquely specified by the node and link
counts, along with a rank-encoded linklist. Also, nodes may link to
themselves, so $L=n^2$. Consequently, their
complexity measure will be
\begin{equation}
{\cal C} = 2\lceil\log_2n\rceil+\lceil\log_2L\rceil+1 + \left\lceil\log_2
\frac{L!}{(L-l)! l!}\right\rceil
\end{equation}
This is, to within a term of order $\log_2 n$ corresponding to
the lead-in prefix specifying the node count, equal to the unnumbered formula
given on page 326 of \cite{Dorin-Korb10}.

\section{New complexity measure compared with the previous proposals}

\begin{figure}
\begin{center}
\resizebox{\textwidth}{!}{\includegraphics{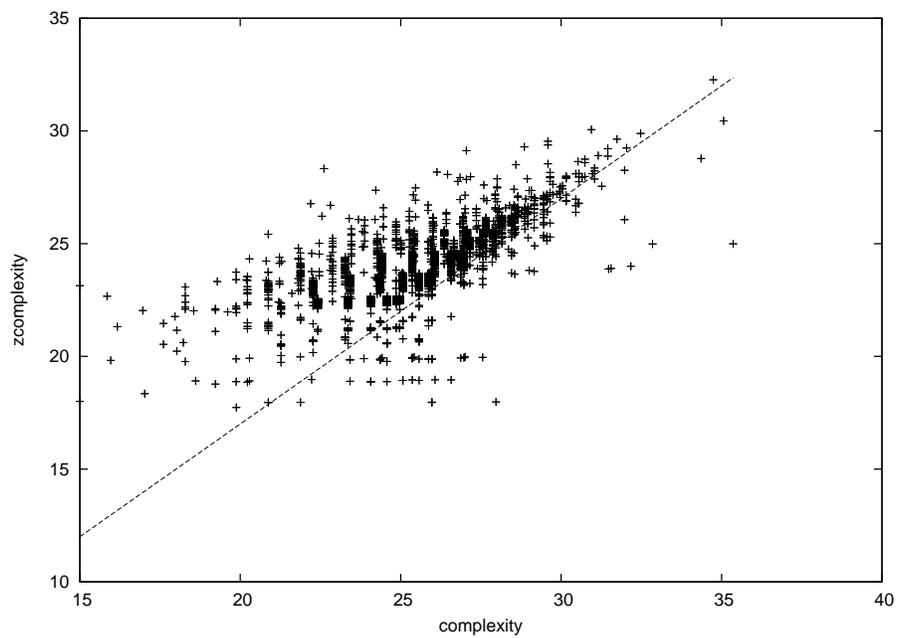}}
\end{center}
\caption{${\cal C}_z$ plotted against 
  ${\cal C}$ for all networks of order 8. The diagonal line
  corresponds to ${\cal C}={\cal C}_z+3$.}
\label{Cz vs newC}
\end{figure}

\begin{figure}
\begin{center}
\resizebox{\textwidth}{!}{\includegraphics{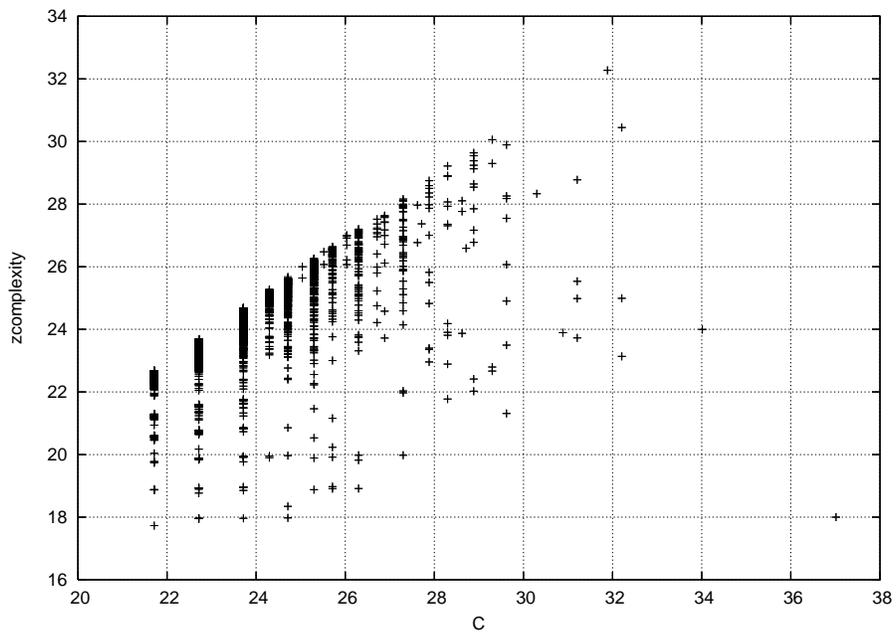}}
\end{center}
\caption{${\cal C}_z$ plotted against the original ${\cal C}$ from
  \cite{Standish05a} for all networks of order 8, reproduced from that
  paper.}
\label{Cz vs C}
\end{figure}

Figures \ref{Cz vs newC} and \ref{Cz vs C} \cite[Fig. 1]{Standish05a}
shows zcomplexity plotted against the new complexity and the original
complexity proposal for all networks of order 8 respectively. The new
complexity proposal is quite well correlated with zcomplexity, and is
in this example about 3 bits higher than zcomplexity. This is just the
difference in prefixes between the two schemes (the original proposal
used 8 1 bits followed by a stop bit = 9 bits overall to represent the node
count, and the new scheme uses 3 1 bits and a stop bit to represent
the field width of the node count, 3 bits for the node field and 5
bits for the link count field = 12 bits overall). Data points
appearing above the line correspond to networks whose linkfields
compress better with the new algorithm than the scheme used with
zcomplexity. Conversely, data points below the line are better
compressed with the old algorithm.

We can conclude that the new scheme usually compresses the link list field
better than the run length encoding scheme employed in
\cite{Standish05a}, and so is a better measure of complexity than
zcomplexity, as well as being far more tractable. The slightly more
complex prefix of the new scheme grows logarithmically with node
count, so will ultimately be more compressed than the prefix of the
old scheme, which grows linearly.

\section{Comparison with medium articulation}

In the last few years Wilhelm\cite{Wilhelm-Hollunder07, Kim-Wilhelm08}
introduced a new complexity like measure that addresses the intuition
that complexity should be minimal for the empty and full networks, and
peak for intermediate values (like figure \ref{l-C}). It is obtained
by multiplying an information quantity that increases with link count
by a different information that falls. The resulting measure is
therefore in square bits, so one should perhaps consider the square
root as the complexity measure.

Precisely, medium articulation is given by
\begin{equation}
\textrm{MA} = -\sum_{ij} w_{ij} \log \frac{w_{ij}}{\sum_k w_{ik}\sum_k
w_{kj}}
        \times \sum_{ij} w_{ij} \log \frac{w_{ij}^2}{\sum_k w_{ik}\sum_k
w_{kj}},
\end{equation}
where $w_{ij}$ is the normalised weight ($\sum_{ij}w_{ij}=1$) 
of the link from node $i$ to node $j$.
 
\begin{figure}
\input{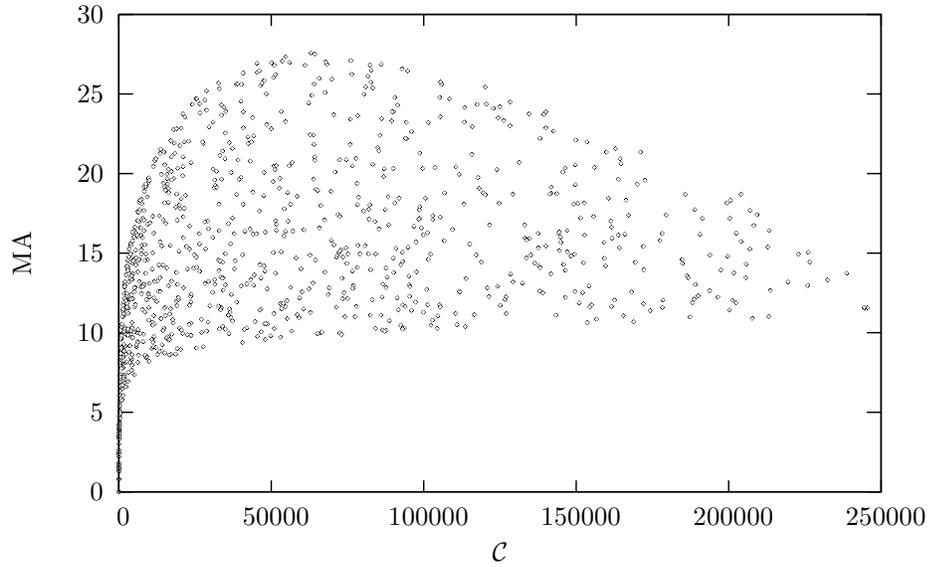}
\caption{Medium Articulation plotted against complexity for 1000
randomly sampled Erd\"os-R\'enyi graphs up to order 500.}
\label{CMA}
\end{figure}

\begin{figure}
\input{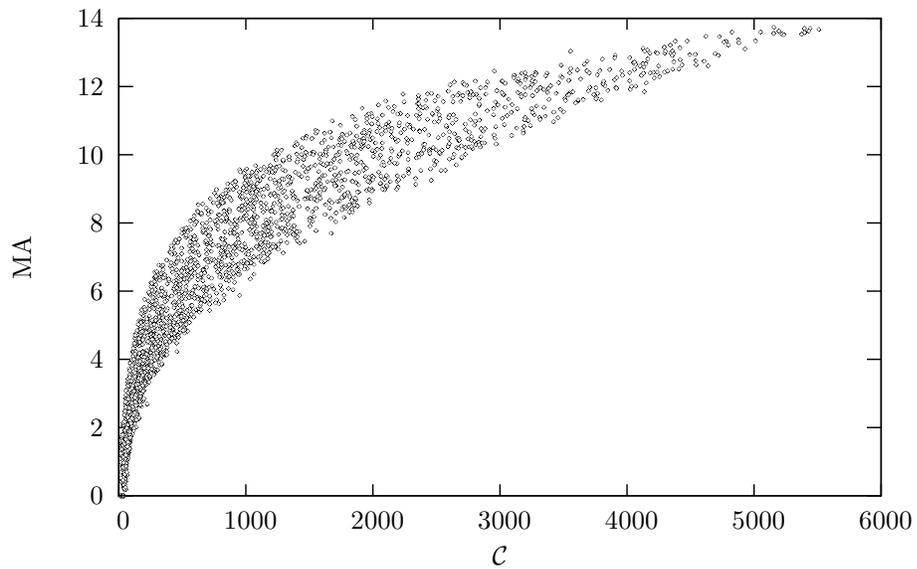}
\caption{Medium Articulation plotted against complexity for 1000
randomly sampled Erd\"os-R\'enyi graphs up to order 500 with no more
than $2n$ links.}
\label{CMA1}
\end{figure}

Figure \ref{CMA} shows medium articulation plotted against ${\cal
C}$ for a sample of 1000 Erd\"os-R\'enyi networks up to order
500. There is no clear relationship between medium articulation and
complexity for the average network. Medium articulation does not
appear to discriminate between complex networks. however if we restrict our
attention to simple networks (Figures \ref{CMA1} and \ref{CMA2})
medium articulation is strongly correlated with complexity, and so
could be used as a proxy for complexity for these cases.

\begin{figure}
\input{cma2}
\caption{Medium Articulation plotted against complexity for 1000
randomly sampled Erd\"os-R\'enyi graphs up to order 500 with more
than $n(n-5)/2$ links.}
\label{CMA2}
\end{figure}

\section{Weighted links}

Whilst the information contained in link weights might be significant
in some circumstances (for instance the weights of a neural network
can only be varied in a limited range without changing the overall
qualitative behaviour of the network), of particular theoretical
interest is to consider the weights as continuous parameters
connecting one network structure with another. For instance if a
network $X$ has the same network structure as the unweighted graph A,
with $b$ links of weight 1 describing the graph $B$ and the
remaining $a-b$ links of weight $w$, then we would like the network
complexity of $X$ to vary smoothly between that of $A$ and $B$ as $w$
varies from 1 to 0. \cite{Gornerup-Crutchfield08} introduced a similar
measure.

The most obvious way of defining this continuous complexity measure is
to start with normalised weights $\sum_iw_i=1$. Then arrange the links
in weight order, and compute the complexity of networks with just
those links of weights less than $w$. The final complexity value is
obtained by integrating:
\begin{equation}\label{weighted C}
{\cal C}(X=N\times L) = \int_0^1 {\cal C}(N\times\{i\in L: w_i<w\}) dw
\end{equation}
Obviously, since the integrand is a stepped function, this is computed
in practice by a sum of complexities of partial networks.

\section{Comparing network complexity with the Erd\"os-R\'enyi random model}

I applied the weighted network complexity measure (\ref{weighted C})
to several well-known real network datasets, obtained from Mark
Newman's website\cite{Watts-Strogatz98,Knuth93,Newman06},
 the Pajek website\cite{Christian-Luczkovich99,Ulanowicz-etal98,Ulanowicz-etal00,Hagy02,Baird-etal98,Almunia-etal99,Baird-Ulanowicz89}
and Duncan Watt's website\cite{Duch-Arenas05,Jeong-etal01},
 with the results
shown in Table \ref{Tab1}. The number of nodes and links of the
networks varied greatly, so the raw complexity values are not
particularly meaningful, as the computed value is highly dependent on
these two network parameters. What is needed is some neutral model for
each network to compare the results to.

At first, one might want to compare the values to an Erd\"os-R\'enyi
random network with the same number of nodes and links. However, in
practice, the real network complexities are much less than that of an
ER network with the same number of nodes and links. This is because in
our scheme, a network with weighted link weights looks somewhat like a
simpler network formed by removing some of the weakest links from the
original. An obvious neutral network model with weighted network
links has the weights drawn from a normal distribution with mean 0.
The sign of the weight can be interpreted as the link direction.
Because the weights in equation (\ref{weighted C}) are normalised, the
complexity value is independent of the standard deviation of the
normal distribution. However, such networks are still much more
complex than the real networks, as the link weight distribution
doesn't match that found in the real network.

\begin{figure}
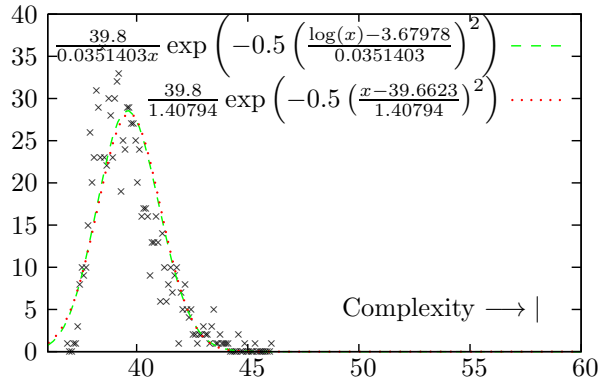

\ifx\PSTloaded\undefined
\def\PSTloaded{t}
\psset{arrowsize=.01 3.2 1.4 .3}
\psset{dotsize=.1}
\catcode`@=11

\newpsobject{PST@Border}{psline}{linewidth=.0015,linestyle=solid}
\newpsobject{PST@Axes}{psline}{linewidth=.0015,linestyle=dotted,dotsep=.004}
\newpsobject{PST@Solid}{psline}{linewidth=.0015,linestyle=solid}
\newpsobject{PST@Dashed}{psline}{linecolor=green,linewidth=.0015,linestyle=dashed,dash=.01 .01}
\newpsobject{PST@Dotted}{psline}{linecolor=red,linewidth=.0025,linestyle=dotted,dotsep=.008}
\newpsobject{PST@LongDash}{psline}{linewidth=.0015,linestyle=dashed,dash=.02 .01}
\newpsobject{PST@Diamond}{psdots}{linewidth=.001,linestyle=solid,dotstyle=square,dotangle=45}
\newpsobject{PST@Filldiamond}{psdots}{linewidth=.001,linestyle=solid,dotstyle=square*,dotangle=45}
\newpsobject{PST@Cross}{psdots}{linewidth=.01,linestyle=solid,dotstyle=+,dotangle=45}
\newpsobject{PST@Plus}{psdots}{linewidth=.001,linestyle=solid,dotstyle=+}
\newpsobject{PST@Square}{psdots}{linewidth=.001,linestyle=solid,dotstyle=square}
\newpsobject{PST@Circle}{psdots}{linewidth=.001,linestyle=solid,dotstyle=o}
\newpsobject{PST@Triangle}{psdots}{linewidth=.001,linestyle=solid,dotstyle=triangle}
\newpsobject{PST@Pentagon}{psdots}{linewidth=.001,linestyle=solid,dotstyle=pentagon}
\newpsobject{PST@Fillsquare}{psdots}{linewidth=.001,linestyle=solid,dotstyle=square*}
\newpsobject{PST@Fillcircle}{psdots}{linewidth=.001,linestyle=solid,dotstyle=*}
\newpsobject{PST@Filltriangle}{psdots}{linewidth=.001,linestyle=solid,dotstyle=triangle*}
\newpsobject{PST@Fillpentagon}{psdots}{linewidth=.001,linestyle=solid,dotstyle=pentagon*}
\newpsobject{PST@Arrow}{psline}{linewidth=.001,linestyle=solid}
\catcode`@=12

\fi
\psset{unit=5.0in,xunit=3.3in,yunit=2in}
\pspicture(0.000000,0.000000)(1.000000,1.000000)
\ifx\nofigs\undefined
\catcode`@=11

\PST@Border(0.1010,0.0840)
(0.1160,0.0840)

\PST@Border(0.9470,0.0840)
(0.9320,0.0840)

\rput[r](0.0850,0.0840){ 0}
\PST@Border(0.1010,0.1945)
(0.1160,0.1945)

\PST@Border(0.9470,0.1945)
(0.9320,0.1945)

\rput[r](0.0850,0.1945){ 5}
\PST@Border(0.1010,0.3050)
(0.1160,0.3050)

\PST@Border(0.9470,0.3050)
(0.9320,0.3050)

\rput[r](0.0850,0.3050){ 10}
\PST@Border(0.1010,0.4155)
(0.1160,0.4155)

\PST@Border(0.9470,0.4155)
(0.9320,0.4155)

\rput[r](0.0850,0.4155){ 15}
\PST@Border(0.1010,0.5260)
(0.1160,0.5260)

\PST@Border(0.9470,0.5260)
(0.9320,0.5260)

\rput[r](0.0850,0.5260){ 20}
\PST@Border(0.1010,0.6365)
(0.1160,0.6365)

\PST@Border(0.9470,0.6365)
(0.9320,0.6365)

\rput[r](0.0850,0.6365){ 25}
\PST@Border(0.1010,0.7470)
(0.1160,0.7470)

\PST@Border(0.9470,0.7470)
(0.9320,0.7470)

\rput[r](0.0850,0.7470){ 30}
\PST@Border(0.1010,0.8575)
(0.1160,0.8575)

\PST@Border(0.9470,0.8575)
(0.9320,0.8575)

\rput[r](0.0850,0.8575){ 35}
\PST@Border(0.1010,0.9680)
(0.1160,0.9680)

\PST@Border(0.9470,0.9680)
(0.9320,0.9680)

\rput[r](0.0850,0.9680){ 40}
\PST@Border(0.2420,0.0840)
(0.2420,0.1040)

\PST@Border(0.2420,0.9680)
(0.2420,0.9480)

\rput(0.2420,0.0420){ 40}
\PST@Border(0.4183,0.0840)
(0.4183,0.1040)

\PST@Border(0.4183,0.9680)
(0.4183,0.9480)

\rput(0.4183,0.0420){ 45}
\PST@Border(0.5945,0.0840)
(0.5945,0.1040)

\PST@Border(0.5945,0.9680)
(0.5945,0.9480)

\rput(0.5945,0.0420){ 50}
\PST@Border(0.7708,0.0840)
(0.7708,0.1040)

\PST@Border(0.7708,0.9680)
(0.7708,0.9480)

\rput(0.7708,0.0420){ 55}
\PST@Border(0.9470,0.0840)
(0.9470,0.1040)

\PST@Border(0.9470,0.9680)
(0.9470,0.9480)

\rput(0.9470,0.0420){ 60}
\PST@Border(0.1010,0.9680)
(0.1010,0.0840)
(0.9470,0.0840)
(0.9470,0.9680)
(0.1010,0.9680)

\rput[r](0.8836,0.1945){Complexity $\longrightarrow|$}
\PST@Cross(0.1325,0.1061)
\PST@Cross(0.1358,0.0840)
\PST@Cross(0.1390,0.0840)
\PST@Cross(0.1423,0.1061)
\PST@Cross(0.1456,0.1061)
\PST@Cross(0.1488,0.1503)
\PST@Cross(0.1521,0.2608)
\PST@Cross(0.1554,0.3050)
\PST@Cross(0.1586,0.2829)
\PST@Cross(0.1619,0.3050)
\PST@Cross(0.1652,0.4155)
\PST@Cross(0.1685,0.6586)
\PST@Cross(0.1717,0.5260)
\PST@Cross(0.1750,0.5923)
\PST@Cross(0.1783,0.7691)
\PST@Cross(0.1815,0.7249)
\PST@Cross(0.1848,0.5923)
\PST@Cross(0.1881,0.8796)
\PST@Cross(0.1913,0.5923)
\PST@Cross(0.1946,0.5702)
\PST@Cross(0.1979,0.7470)
\PST@Cross(0.2011,0.7028)
\PST@Cross(0.2044,0.5923)
\PST@Cross(0.2077,0.7470)
\PST@Cross(0.2110,0.7912)
\PST@Cross(0.2142,0.8133)
\PST@Cross(0.2175,0.5039)
\PST@Cross(0.2208,0.6365)
\PST@Cross(0.2240,0.6144)
\PST@Cross(0.2273,0.7249)
\PST@Cross(0.2306,0.7249)
\PST@Cross(0.2338,0.6807)
\PST@Cross(0.2371,0.6807)
\PST@Cross(0.2404,0.6365)
\PST@Cross(0.2436,0.5260)
\PST@Cross(0.2469,0.6144)
\PST@Cross(0.2502,0.4376)
\PST@Cross(0.2534,0.4597)
\PST@Cross(0.2567,0.4597)
\PST@Cross(0.2600,0.4376)
\PST@Cross(0.2633,0.2829)
\PST@Cross(0.2665,0.3713)
\PST@Cross(0.2698,0.3713)
\PST@Cross(0.2731,0.4376)
\PST@Cross(0.2763,0.3713)
\PST@Cross(0.2796,0.2166)
\PST@Cross(0.2829,0.3934)
\PST@Cross(0.2861,0.3050)
\PST@Cross(0.2894,0.2166)
\PST@Cross(0.2927,0.2608)
\PST@Cross(0.2959,0.3050)
\PST@Cross(0.2992,0.2387)
\PST@Cross(0.3025,0.2829)
\PST@Cross(0.3058,0.3050)
\PST@Cross(0.3090,0.1061)
\PST@Cross(0.3123,0.2387)
\PST@Cross(0.3156,0.1945)
\PST@Cross(0.3188,0.2608)
\PST@Cross(0.3221,0.1945)
\PST@Cross(0.3254,0.1724)
\PST@Cross(0.3286,0.1945)
\PST@Cross(0.3319,0.1282)
\PST@Cross(0.3352,0.1282)
\PST@Cross(0.3384,0.1061)
\PST@Cross(0.3417,0.1282)
\PST@Cross(0.3450,0.1282)
\PST@Cross(0.3483,0.1503)
\PST@Cross(0.3515,0.1282)
\PST@Cross(0.3548,0.1282)
\PST@Cross(0.3581,0.1061)
\PST@Cross(0.3613,0.1503)
\PST@Cross(0.3646,0.1945)
\PST@Cross(0.3679,0.1061)
\PST@Cross(0.3711,0.1061)
\PST@Cross(0.3744,0.1282)
\PST@Cross(0.3777,0.1061)
\PST@Cross(0.3809,0.1061)
\PST@Cross(0.3842,0.1061)
\PST@Cross(0.3875,0.1061)
\PST@Cross(0.3908,0.0840)
\PST@Cross(0.3940,0.0840)
\PST@Cross(0.3973,0.0840)
\PST@Cross(0.4006,0.0840)
\PST@Cross(0.4038,0.1282)
\PST@Cross(0.4071,0.0840)
\PST@Cross(0.4104,0.0840)
\PST@Cross(0.4136,0.1061)
\PST@Cross(0.4169,0.0840)
\PST@Cross(0.4202,0.0840)
\PST@Cross(0.4234,0.0840)
\PST@Cross(0.4267,0.0840)
\PST@Cross(0.4300,0.1061)
\PST@Cross(0.4332,0.0840)
\PST@Cross(0.4365,0.0840)
\PST@Cross(0.4398,0.0840)
\PST@Cross(0.4431,0.0840)
\PST@Cross(0.4463,0.0840)
\PST@Cross(0.4496,0.0840)
\PST@Cross(0.4529,0.0840)
\PST@Cross(0.4561,0.1061)

\rput[r](0.8200,0.8850){$\frac{39.8}{0.0351403x}\exp\left(-0.5\left(\frac{\log(x)-3.67978}{0.0351403}\right)^2\right)$}
\PST@Dashed(0.8360,0.8850)
(0.9150,0.8850)

\PST@Dashed(0.1010,0.1003)
(0.1010,0.1003)
(0.1095,0.1109)
(0.1181,0.1265)
(0.1266,0.1487)
(0.1352,0.1788)
(0.1437,0.2178)
(0.1523,0.2662)
(0.1608,0.3233)
(0.1694,0.3875)
(0.1779,0.4559)
(0.1865,0.5244)
(0.1950,0.5886)
(0.2035,0.6436)
(0.2121,0.6851)
(0.2206,0.7097)
(0.2292,0.7155)
(0.2377,0.7024)
(0.2463,0.6718)
(0.2548,0.6267)
(0.2634,0.5709)
(0.2719,0.5087)
(0.2805,0.4444)
(0.2890,0.3816)
(0.2975,0.3233)
(0.3061,0.2714)
(0.3146,0.2271)
(0.3232,0.1905)
(0.3317,0.1613)
(0.3403,0.1388)
(0.3488,0.1219)
(0.3574,0.1096)
(0.3659,0.1009)
(0.3745,0.0949)
(0.3830,0.0909)
(0.3915,0.0883)
(0.4001,0.0866)
(0.4086,0.0855)
(0.4172,0.0849)
(0.4257,0.0845)
(0.4343,0.0843)
(0.4428,0.0842)
(0.4514,0.0841)
(0.4599,0.0840)
(0.4685,0.0840)
(0.4770,0.0840)
(0.4855,0.0840)
(0.4941,0.0840)
(0.5026,0.0840)
(0.5112,0.0840)
(0.5197,0.0840)
(0.5283,0.0840)
(0.5368,0.0840)
(0.5454,0.0840)
(0.5539,0.0840)
(0.5625,0.0840)
(0.5710,0.0840)
(0.5795,0.0840)
(0.5881,0.0840)
(0.5966,0.0840)
(0.6052,0.0840)
(0.6137,0.0840)
(0.6223,0.0840)
(0.6308,0.0840)
(0.6394,0.0840)
(0.6479,0.0840)
(0.6565,0.0840)
(0.6650,0.0840)
(0.6735,0.0840)
(0.6821,0.0840)
(0.6906,0.0840)
(0.6992,0.0840)
(0.7077,0.0840)
(0.7163,0.0840)
(0.7248,0.0840)
(0.7334,0.0840)
(0.7419,0.0840)
(0.7505,0.0840)
(0.7590,0.0840)
(0.7675,0.0840)
(0.7761,0.0840)
(0.7846,0.0840)
(0.7932,0.0840)
(0.8017,0.0840)
(0.8103,0.0840)
(0.8188,0.0840)
(0.8274,0.0840)
(0.8359,0.0840)
(0.8445,0.0840)
(0.8530,0.0840)
(0.8615,0.0840)
(0.8701,0.0840)
(0.8786,0.0840)
(0.8872,0.0840)
(0.8957,0.0840)
(0.9043,0.0840)
(0.9128,0.0840)
(0.9214,0.0840)
(0.9299,0.0840)
(0.9385,0.0840)
(0.9470,0.0840)

\rput[r](0.8200,0.7430){$\frac{39.8}{1.40794}\exp\left(-0.5\left(\frac{x-39.6623}{1.40794}\right)^2\right)$}
\PST@Dotted(0.8360,0.7430)(0.9150,0.7430)

\PST@Dotted(0.1010,0.1052)
(0.1010,0.1052)
(0.1095,0.1167)
(0.1181,0.1329)
(0.1266,0.1551)
(0.1352,0.1843)
(0.1437,0.2214)
(0.1523,0.2667)
(0.1608,0.3198)
(0.1694,0.3795)
(0.1779,0.4434)
(0.1865,0.5084)
(0.1950,0.5705)
(0.2035,0.6254)
(0.2121,0.6689)
(0.2206,0.6975)
(0.2292,0.7086)
(0.2377,0.7014)
(0.2463,0.6764)
(0.2548,0.6358)
(0.2634,0.5830)
(0.2719,0.5221)
(0.2805,0.4574)
(0.2890,0.3929)
(0.2975,0.3321)
(0.3061,0.2774)
(0.3146,0.2304)
(0.3232,0.1916)
(0.3317,0.1608)
(0.3403,0.1372)
(0.3488,0.1197)
(0.3574,0.1073)
(0.3659,0.0988)
(0.3745,0.0931)
(0.3830,0.0894)
(0.3915,0.0871)
(0.4001,0.0858)
(0.4086,0.0850)
(0.4172,0.0845)
(0.4257,0.0843)
(0.4343,0.0841)
(0.4428,0.0841)
(0.4514,0.0840)
(0.4599,0.0840)
(0.4685,0.0840)
(0.4770,0.0840)
(0.4855,0.0840)
(0.4941,0.0840)
(0.5026,0.0840)
(0.5112,0.0840)
(0.5197,0.0840)
(0.5283,0.0840)
(0.5368,0.0840)
(0.5454,0.0840)
(0.5539,0.0840)
(0.5625,0.0840)
(0.5710,0.0840)
(0.5795,0.0840)
(0.5881,0.0840)
(0.5966,0.0840)
(0.6052,0.0840)
(0.6137,0.0840)
(0.6223,0.0840)
(0.6308,0.0840)
(0.6394,0.0840)
(0.6479,0.0840)
(0.6565,0.0840)
(0.6650,0.0840)
(0.6735,0.0840)
(0.6821,0.0840)
(0.6906,0.0840)
(0.6992,0.0840)
(0.7077,0.0840)
(0.7163,0.0840)
(0.7248,0.0840)
(0.7334,0.0840)
(0.7419,0.0840)
(0.7505,0.0840)
(0.7590,0.0840)
(0.7675,0.0840)
(0.7761,0.0840)
(0.7846,0.0840)
(0.7932,0.0840)
(0.8017,0.0840)
(0.8103,0.0840)
(0.8188,0.0840)
(0.8274,0.0840)
(0.8359,0.0840)
(0.8445,0.0840)
(0.8530,0.0840)
(0.8615,0.0840)
(0.8701,0.0840)
(0.8786,0.0840)
(0.8872,0.0840)
(0.8957,0.0840)
(0.9043,0.0840)
(0.9128,0.0840)
(0.9214,0.0840)
(0.9299,0.0840)
(0.9385,0.0840)
(0.9470,0.0840)

\PST@Border(0.1010,0.9680)
(0.1010,0.0840)
(0.9470,0.0840)
(0.9470,0.9680)
(0.1010,0.9680)

\catcode`@=12
\fi
\endpspicture
\caption{The distribution of complexities of the shuffled Narragansett Bay
  food web\cite{Monaco-Ulanowicz97}. Both a normal and a log-normal
  distribution have been fitted to the data --- the log-normal is a
  slightly better fit than the normal distribution. In the bottom
  right hand corner is marked the complexity value computed for the
  actual Narragansett food web (58.2 bits). }
\label{shuffled-foodweb}
\end{figure}

Instead, a simple way of generating a neutral model is to break and
reattach the network links to random nodes, without merging links,
leaving the original link weights
unaltered. In the following experiment, we generate 1000 of these
shuffled networks. The distribution of complexities can be fitted
to a lognormal distribution, which gives a better likelihood than a
normal distribution for all networks studied
here\cite{Clauset-etal07}, although the difference between a
log-normal and a normal fit becomes less pronounced for more complex
networks. Figure \ref{shuffled-foodweb} shows the distribution of
complexity values computed by shuffling the Narragansett food web, and the
best fit normal and lognormal distributions.

In what follows we compute the average $\langle\ln{\cal
  C}_\mathrm{ER}\rangle$ and standard deviation $\sigma_\mathrm{ER}$
  of the {\em logarithm} of the neutral model complexities. We can
  then compare the network complexity with the ensemble of neutral
  models to obtain a $p$-value, the probability that the observed
  complexity is that of a random network. The $p$-values are so small
  that is better to represent the value as a number of standard
  deviations (``sigmas'') that the logarithm of the measured
  complexity value is from the mean logarithm of the shuffled
  networks.  A value of 6 sigmas corresponds to a $p$-value less than
  $1.3\times10^{-4}$, although this must be taken with a certain grain
  of salt, as the distribution of shuffled complexity values has a
  fatter tail than the fitted log-normal distribution. In none of
  these samples did the shuffled network complexity exceed the
  original network's complexity, meaning the $p$-value is less than
  $10^{-3}$. The difference ${\cal C}-\exp(\langle\ln{\cal
  C}_\mathrm{ER}\rangle)$ is the amount of information contained in
  the specific arrangement of links.

A code implementing this algorithm is implemented as a C++ library, and
is available from version 4.D36 onwards as part of the \EcoLab{}
system, an open source modelling framework hosted at
http://ecolab.sourceforge.net.

\section{ALife models}

\subsection{Tierra}\label{tierra}

Tierra \cite{Ray91} is a well known artificial life system in which
self reproducing computer programs written in an assembly-like
language are allowed to evolve. The programs, or {\em digital
  organisms} can interact with each other via template matching operations,
modelled loosely on the way proteins interact in real biological
systems. A number of distinct strategies evolve, including parasitism,
where organisms make use of another organism's code and
hyper-parasitism where an organism sets traps for parasites in order
to steal their CPU resources. At any point in time in a Tierra run,
there is an interaction network between the species present, which is
the closest thing in the Tierra world to a foodweb.

Tierra is an aging platform, with the last release (v6.02) having been
released more than six years ago. For this work, I used an even older
release (5.0), for which I have had some experience in working
with. Tierra was originally written in C for an environment where ints
were 16 bits and long ints 32 bits. This posed a problem for using it
on the current generation of 64 bit computers, where the word sizes
are doubled. Some effort was needed to get the code 64 bit
clean. Secondly a means of extracting the interaction network was
needed. Whilst Tierra provided the concept of ``watch bits'', which
recorded whether a digital organism had accessed another's genome or
vice versa, it did not record which other genome was accessed. So I
modified the template matching code to log the pair of genome
labels that performed the template match to a file.

Having a record of interactions by genotype label, it is necessary to map the
genotype to phenotype. In Tierra, the phenotype is the behaviour of
the digital organism, and can be judged by running the organisms
pairwise in a tournament, to see what effect each has on the
other. The precise details for how this can be done is described in
\cite{Standish03a}.  

Having a record of interactions between phenotypes, and discarding
self-self interactions, there are a number of ways of turning that
record into a foodweb. The simplest way, which I adopted, was sum the
interactions between each pair of phenotypes over a sliding window of
20 million executed instructions, and doing this every 100 million executed
instructions. This lead to time series of around 1000 foodwebs for
each Tierra run. 

In Tierra, parsimony pressure is controlled by the parameter
SlicePow. CPU time is allocated proportional to genome size raised to
SlicePow. If SlicePow is close to 0, then there is great evolutionary
pressure for the organisms to get as small as possible to increase
their replication rate. When it is one, this pressure is
eliminated. In \cite{Standish04c}, I found that a SlicePow of around
0.95 was optimal. If it were much higher, the organisms grow so large and so
rapidly that they eventually occupy more than 50\% of the soup. At
which point they kill the soup at their next Mal (memory allocation)
operation. In this work, I altered the implementation of Mal to fail
if the request was more than than the soup size divided by minimum
population save threshold (usually around 10). Organisms any larger
than this will never appear in the Genebanker (Tierra's equivalent of
the fossil record), as their population can never exceed the save
threshold. This modification allows SlicePow = 1 runs to run for
an extensive period of time without the soup dying. 

\subsection{EcoLab}

EcoLab was introduced by the author as a simple model of an evolving
ecosystem \cite{Standish94}. The ecological dynamics is described by an
$n$-dimensional generalised Lotka-Volterra equation:
\begin{equation}
\dot{n_i} = r_in_i + \sum_j\beta_{ij}n_in_j,
\end{equation}
where $n_i$ is the population density of species $i$, $r_i$ its growth
rate and $\beta_{ij}$ the interaction matrix. Extinction is handled
via a novel stochastic truncation algorithm, rather than the more usual
threshold method. Speciation occurs by randomly mutating the ecological
parameters ($r_i$ and $\beta_{ij}$) of the parents, subject to the
constraint that the system remain bounded \cite{Standish98b}.

The interaction matrix is a candidate foodweb, but has too much
information. Its offdiagonal terms may be negative as well as
positive, whereas for the complexity definition (\ref{weighted C}), we
need the link weights to be positive. There are a number of ways of
resolving this issue, such as ignoring the sign of the off-diagonal
term (ie taking its absolute value), or antisymmetrising the matrix by
subtracting its transpose, then using the sign of the offdiagonal term
to determine the link direction.

For the purposes of this study, I chose to subtract just the negative
$\beta_{ij}$ terms from themselves and their transpose terms $\beta_{ji}$.
This effects a maximal encoding of the interaction matrix information in
the network structure, with link direction and weight encoding the
direction and size of resource flow. The effect is as follows:
\begin{itemize}
\item Both $\beta_{ij}$ and $\beta_{ji}$ are positive (the {\em
    mutualist} case). Neither offdiagonal term changes, and the two nodes
    have links pointing in both directions, with weights given by the
    two offdiagonal terms.
\item Both $\beta_{ij}$ and $\beta_{ji}$ are negative (the {\em
    competitive} case). The terms are swapped, and the signs changed
    to be positive. Again the two nodes have links pointing in both
    directions, but the link direction reflects the direction of
    resource flow.
\item Both $\beta_{ij}$ and $\beta_{ji}$ are of opposite sign (the
  {\em predator-prey} or {\em parasitic} case). Only a single link
  exists between species $i$ and $j$, whose weight is the summed
  absolute values of the offdiagonal terms, and whose link direction
  reflects the direction of resource flow.
\end{itemize}

\subsection{Webworld}

Webworld is another evolving ecology model, similar in some respects
to EcoLab, introduced by \cite{Caldarelli-etal98}, with some
modifications described in \cite{Drossel-etal01}. It features more
realistic ecological interactions than does EcoLab, in that it tracks
biomass resources. It too has an interaction matrix called a {\em
  functional response} in that model that could serve as a foodweb,
which is converted to a directed weighted graph in the same way as the
EcoLab interaction matrix. I used the Webworld implementation
distributed with the \EcoLab{} simulation platform
\cite{Standish04a}. The parameters were chosen as $R=10^5$,
$b=5\times10^{-3}$, $c=0.8$ and $\lambda=0.1$.

\section{Methods and materials}

Tierra was run on a 512KB soup, with SlicePow set to 1, until the soup
died, typically after some $5\times10^{10}$ instructions have executed. Some
variant runs were performed with SlicePow=0.95, and with different random
number generators, but no difference in the outcome was observed.

The source code of Tierra 5.0 was modified in a few places, as
described in the Tierra section of this paper. The final source code
is available as tierra.5.0.D7.tar.gz from the \EcoLab{} website hosted
on SourceForge (http://ecolab.sf.net).

The genebanker output was processed by the eco-tierra.3.D13 code, also
available from the \EcoLab{} website, to produce a list of phenotype
equivalents for each genotype. A function for processing the
interaction log file generated by Tierra and producing a timeseries of
foodweb graphs was added to Eco-tierra. The script for running this
postprocessing step is process\_ecollog.tcl.

The EcoLab model was adapted to convert the interaction matrix into a
foodweb and log the foodweb to disk every 1000 time steps for later
processing. The Webworld model was adapted similarly. The model parameters
were as documented in the included ecolab.tcl and webworld.tcl experiment files
of the ecolab.5.D1 distribution, which is also available from the \EcoLab{}
website.

Finally, each foodweb, whether real world, or ALife generated, and 100
link-shuffled control versions were run through the network complexity
algorithm (\ref{weighted C}). This is documented in the cmpERmodel.tcl
script of ecolab.5.D1.  The average and standard deviation of $\ln
{\cal C}$ was calculated, rather than ${\cal C}$ directly, as the
shuffled complexity values fitted a log-normal distribution better
than a standard normal distribution.  The difference between the
measured complexity and $\exp\langle\ln {\cal C}\rangle$ (ie the
geometric mean of the control network complexities) is what is
reported as the surplus in Figures \ref{start-fig}--\ref{end-fig}.

\section{ALife results}

\begin{figure*}
\small
\begin{center}
\resizebox{\textwidth}{!}{\includegraphics{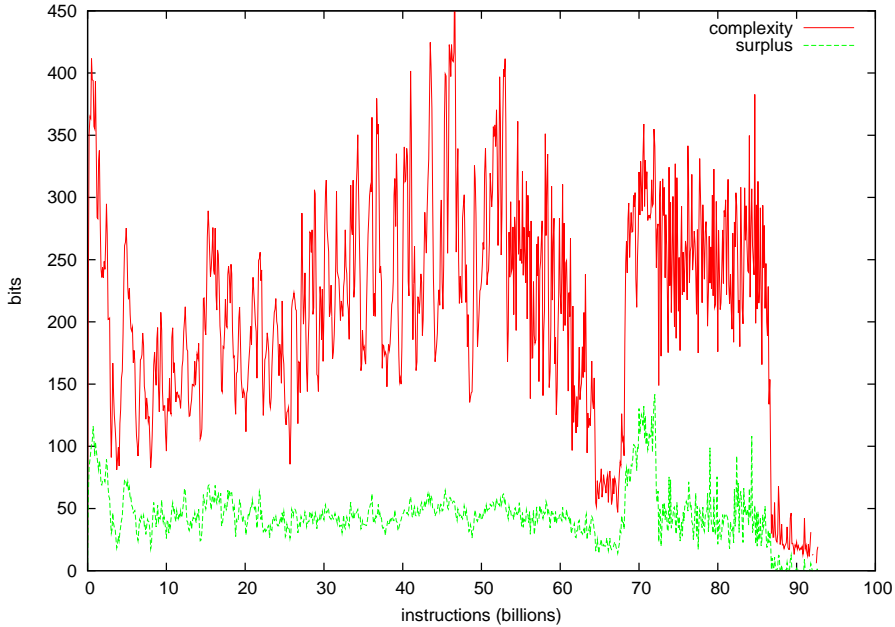}}\\
\end{center}
\caption{Complexity of the Tierran interaction network for
  SlicePow=0.95, and the associated complexity surplus. The surplus
  was typically 15--30 times the standard deviation of the ER neutral
  network complexities.}
\label{start-fig}\label{tierra-results}
\end{figure*}

\begin{figure*}
\small
\begin{center}
\resizebox{\textwidth}{!}{\includegraphics{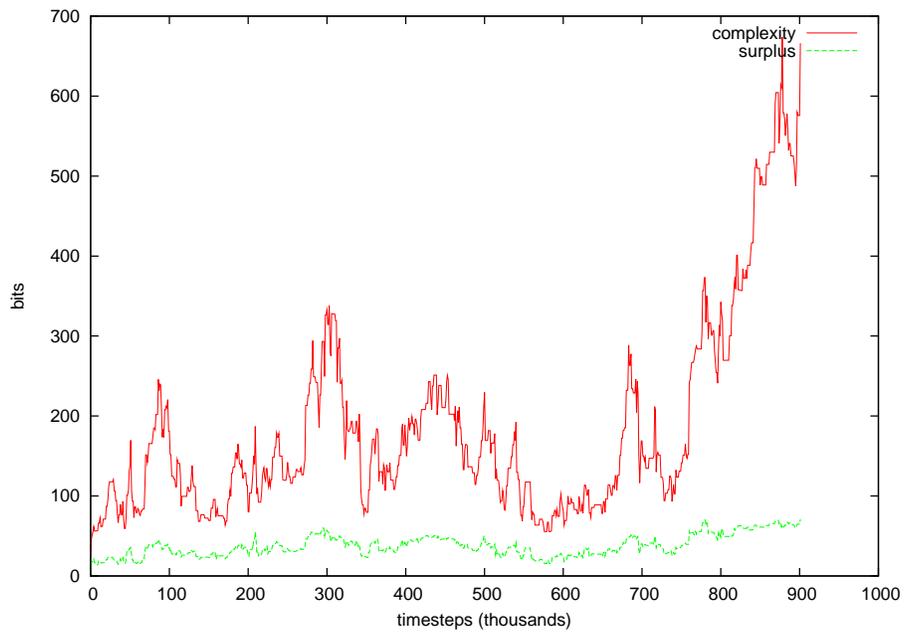}}\\
\end{center}
\caption{Complexity of EcoLab's foodweb, and the associated complexity
surplus. For most of the run, the surplus
  was  10--40 times the standard deviation of the ER neutral
  network complexities, but in the final complexity growth phase, it
  was several 100.}
\end{figure*}

\begin{figure*}
\begin{center}
\resizebox{\textwidth}{!}{\includegraphics{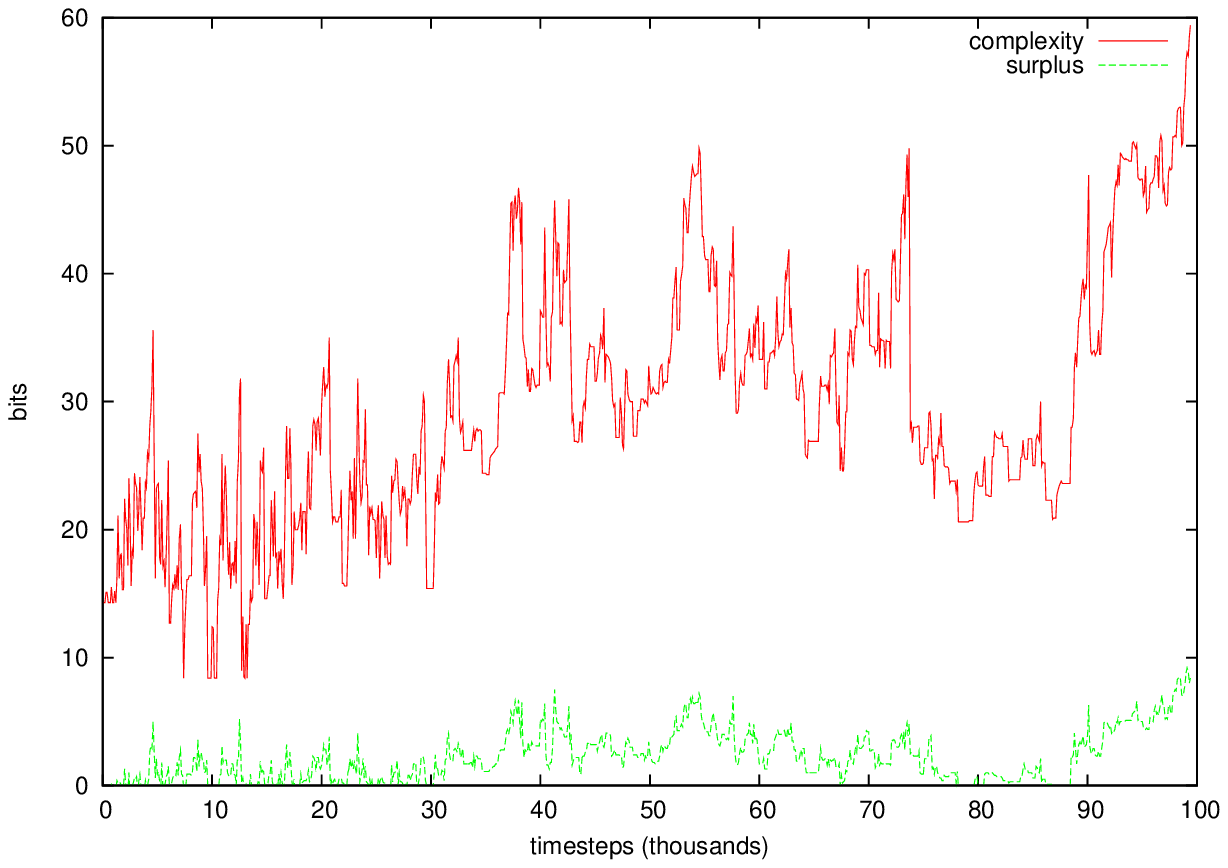}}
\end{center}
\caption{Complexity of Webworld's foodweb, and the associated complexity
surplus. The surplus was typically 1--5 times the standard deviation of the ER neutral network complexities.}
\label{end-fig}
\end{figure*}

Figures \ref{start-fig}--\ref{end-fig} show the computed complexity
values and the surplus of the complexity over the average randomly
shuffled networks. In both the EcoLab and Tierra cases, significant
amounts of surplus complexity can be observed, but not in the case of
WebWorld. In WebWorld's case, this is probably because the system
complexity was not very high in the first place --- the diversity
typically ranged between 5 and 10 species throughout the WebWorld
run. It is possible that the competition parameter $c$ was too high
for this run to generate any appreciable foodweb complexity.

An earlier report of these results \cite{Standish10b} did not show
significant complexity, unlike what is shown here. It was discovered
that the process of writing the foodweb data to an intermediate file
stripped the weight data from the network links, preserving only the
graph's topological structure. It turns out that the complexity surplus
phenomenon is only present in weighted networks --- unweighted
networks tend to be indistinguishable from randomly generated networks.

\section{Discussion}

In this paper, a modified version of a previously proposed simple
representation language for $n$-node directed graphs is given. This
modification leads to a more intuitive complexity measure that gives a
minimum value for empty and full networks. When compared with the
zcomplexity measure introduced in the previous paper, it is somewhat
correlated with, and is at least as good a measure as zcomplexity, but
has the advantage of being far more tractable. 

When compared with random networks created by shuffling the links, all
real world example networks exhibited higher complexity than the
random network. The difference between the complexity of the real
network and the mean of the complexities of the randomly shuffled
network represents structural information contained in the specific
arrangement of links making up the network. To test the hypothesis
that networks generated by evolutionary selection will also have a
complexity surplus representing the functional nature of the network,
I ran the same analysis on foodwebs generated by various artificial
life systems. These too, exhibited the same complexity surplus seen
in the real world networks. 

By contrast, I have applied this technique to some scale free networks
generated by the Barab\'asi-Albert {\em preferential attachment
process}\cite{Barabasi-Albert99}, with varying numbers of attachment
points per node and different weight distributions. On none of these
graphs were significant complexity surpluses generated.

If the weights are stripped off the edges, then the effect disappears,
which provides a clue as to the origin of the effect. In equation
(\ref{weighted C}), the components dominating the sum will tend to be
connected giant components of the overall network with very few
cycles. The randomly shuffled versions of these, however, will tend to
be composed of small disconnected components, with many of the same
motifs represented over and over again. Consequently, the shuffled
network has a great deal of symmetry, reducing the overall
complexity. Even though naively one might think that random networks
should be the most complex within the class of networks of a given
node and link count, given the nature of Kolmogorov complexity to
favour random strings, the weighted complexity formula (\ref{weighted
C}) is sensitive to the structure within the network that has meaning
to system behaviour that the network represents.

Reverting to an earlier posed question about the amount of complexity
stored within the foodweb as compared with individual organism's
phenotypic complexity, the results in Figure \ref{tierra-results}
gives a preliminary answer. Each foodweb in Figure
\ref{tierra-results} averaged around 100 species, the average
phenotypic complexity of which is around 150--200 bits (750 bits at
maximum)\cite{Standish03a}. So the 250--400 bits of network complexity
is but a drop in the bucket of phenotypic complexity.

\bibliographystyle{alj}
\bibliography{rus}

\begin{thebibliography}{10}

\bibitem{Almunia-etal99}
Almunia, J., Basterretxea, G., Aristegui, J., and Ulanowicz., R. (1999)
  Benthic- pelagic switching in a coastal subtropical lagoon. {\em Estuarine,
  Coastal and Shelf Science\/}, {\bf 49}, 363--384.

\bibitem{Baird-etal98}
Baird, D., Luczkovich, J., and Christian, R.~R. (1998) Assessment of spatial
  and temporal variability in ecosystem attributes of the {St Marks National
  Wildlife Refuge}, {Apalachee Bay, Florida}. {\em Estuarine, Coastal, and
  Shelf Science\/}, {\bf 47}, 329--349.

\bibitem{Baird-Ulanowicz89}
Baird, D. and Ulanowicz, R. (1989) The seasonal dynamics of the {Chesapeake
  Bay} ecosystem. {\em Ecological Monographs\/}, {\bf 59}, 329--364.

\bibitem{Barabasi-Albert99}
Barab\'asi, A.-L. and Albert, R. (1999) Emergence of scaling in random
  networks. {\em Science\/}, {\bf 286}, 509--512.

\bibitem{Bedau-etal00}
Bedau, M.~A., McCaskill, J.~S., Packard, N.~H., Rasmussen, S., Adami, C.,
  Green, D.~G., Ikegami, T., Kaneko, K., and Ray, T.~S. (2000) Open problems in
  artificial life. {\em Artificial Life\/}, {\bf 6}, 363--376.

\bibitem{Caldarelli-etal98}
Caldarelli, G., Higgs, P.~G., and McKane, A.~J. (1998) Modelling coevolution in
  multispecies communities. {\em J. Theor, Biol.\/}, {\bf 193}, 345--358.

\bibitem{Christian-Luczkovich99}
Christian, R. and Luczkovich, J. (1999) Organizing and understanding a winter's
  seagrass foodweb network through effective trophic levels. {\em Ecological
  Modelling\/}, {\bf 117}, 99--124.

\bibitem{Clauset-etal07}
Clauset, A., Shalizi, C.~R., and Newman, M. E.~J. (2009) Power-law
  distributions in empirical data. {\em SIAM Review\/}, {\bf 51}, 661--703,
  arXiv:0706.1062.

\bibitem{Claussen04}
Claussen, J.~C. (2007) Offdiagonal complexity: A computationally quick
  complexity measure for graphs and networks. {\em Physica A\/}, {\bf 375},
  365--373, arXiv:q-bio.MN/0410024.

\bibitem{Dewar03}
Dewar, R. (2003) Informational theory explanation of the fluctuation theorem,
  maximum entropy production and self-organized criticality in non-equilibrium
  stationary states. {\em J. Phys. A\/}, {\bf 36}, 631--641.

\bibitem{Dorin-Korb10}
Dorin, A. and Korb, K.~B. (2010) Network measures of ecosystem complexity.
  Fellerman et~al.  \cite{ALifeXII}, pp. 323--328.

\bibitem{Drossel-etal01}
Drossel, B., Higgs, P.~G., and McKane, A.~J. (2001) The influence of
  predator-prey population dynamics on the long-term evolution of food web
  structure. {\em J. Theor. Biol.\/}, {\bf 208}, 91--107.

\bibitem{Duch-Arenas05}
Duch, J. and Arenas, A. (2005) Community identification using extremal
  optimization. {\em Physical Review E\/}, {\bf 72}, 027104.

\bibitem{Edmonds99}
Edmonds, B. (1999) {\em Syntactic Measures of Complexity\/}. Ph.D. thesis,
  University of Manchester, http://bruce.edmonds.name/thesis/, Complexity
  Bibliography: http://bruce.edmonds.name/compbib/.

\bibitem{ALifeXII}
Fellerman, H. et~al. (eds.) (2010) {\em Proceedings of Artificial Life XII\/},
  MIT Press.

\bibitem{Gornerup-Crutchfield08}
G\"ornerup, O. and Crutchfield, J.~P. (2008) Hierarchical self-organization in
  the finitary process soup. {\em Artificial Life\/}, {\bf 14}, 245--254.

\bibitem{Hagy02}
Hagy, J. (2002) {\em Eutrophication, hypoxia and trophic transfer efficiency in
  {Chesapeake Bay}\/}. Ph.D. thesis, University of Maryland at College Park
  (USA).

\bibitem{Jeong-etal01}
Jeong, H., Mason, S., Barab\'asi, A.-L., and Oltvai, Z.~N. (2001) Centrality
  and lethality of protein networks. {\em Nature\/}, {\bf 411}, 41.

\bibitem{Kim-Wilhelm08}
Kim, J. and Wilhelm, T. (2008) What is a complex graph. {\em Physica A\/}, {\bf
  387}, 2637--2652.

\bibitem{Knuth93}
Knuth, D.~E. (1993) {\em The {Stanford GraphBase}: A Platform for Combinatorial
  Computing\/}. Addison-Wesley.

\bibitem{Layzer88}
Layzer, D. (1988) Growth and order in the universe. Weber, B.~H., Depew, D.~J.,
  and Smith, J.~D. (eds.), {\em Entropy, Information and Evolution\/}, pp.
  23--39, MIT Press.

\bibitem{Li-Vitanyi97}
Li, M. and Vit\'anyi, P. (1997) {\em An Introduction to {Kolmogorov} Complexity
  and its Applications\/}. Springer, 2nd edn.

\bibitem{McKay81}
McKay, B.~D. (1981) Practical graph isomorphism. {\em Congressus
  Numerantium\/}, {\bf 30}, 45--87.

\bibitem{Monaco-Ulanowicz97}
Monaco, M. and Ulanowicz, R. (1997) Comparative ecosystem trophic structure of
  three {U.S. Mid-Atlantic} estuaries. {\em Mar. Ecol. Prog. Ser.\/}, {\bf
  161}, 239--254.

\bibitem{Myrvold-Ruskey01}
Myrvold, W. and Ruskey, F. (2001) Ranking and unranking permutations in linear
  time. {\em Information Processing Letters\/}, {\bf 79}, 281--284.

\bibitem{Newman06}
Newman, M. E.~J. (2006) Finding community structure in networks using the
  eigenvectors of matrices. {\em Phys. Rev. E\/}, {\bf 74}, 036104.

\bibitem{Prigogine80}
Prigogine, I. (1980) {\em From Being into Becoming\/}. W.H. Freeman.

\bibitem{Prokopenko-etal09}
Prokopenko, M., Boschetti, F., and Ryan, A.~J. (2009) An information-theoretic
  primer on complexity, self-organization, and emergence. {\em Complexity\/},
  {\bf 15}, 11--28.

\bibitem{Ray91}
Ray, T. (1991) An approach to the synthesis of life. Langton, C.~G., Taylor,
  C., Farmer, J.~D., and Rasmussen, S. (eds.), {\em Artificial Life {II}\/}, p.
  371, Addison-Wesley.

\bibitem{Standish98b}
Standish, R.~K. (2000) The role of innovation within economics. Barnett, W.,
  Chiarella, C., Keen, S., Marks, R., and Schnabl, H. (eds.), {\em Commerce,
  Complexity and Evolution\/}, vol.~11 of {\em International Symposia in
  Economic Theory and Econometrics\/}, pp. 61--79, Cambridge UP.

\bibitem{Standish94}
Standish, R.~K. (1994) Population models with random embryologies as a paradigm
  for evolution. {\em Complexity International\/}, {\bf 2}.

\bibitem{Standish01a}
Standish, R.~K. (2001) On complexity and emergence. {\em Complexity
  International\/}, {\bf 9}, arXiv:nlin.AO/0101006.

\bibitem{Standish03a}
Standish, R.~K. (2003) Open-ended artificial evolution. {\em International
  Journal of Computational Intelligence and Applications\/}, {\bf 3}, 167,
  arXiv:nlin.AO/0210027.

\bibitem{Standish04a}
Standish, R.~K. (2004) {Ecolab}, {Webworld} and self-organisation. Pollack
  et~al. (eds.), {\em Artificial Life IX\/}, Cambridge, MA, p. 358, MIT Press.

\bibitem{Standish04c}
Standish, R.~K. (2004) The influence of parsimony and randomness on complexity
  growth in {Tierra}. Bedau et~al. (eds.), {\em ALife IX Workshop and Tutorial
  Proceedings\/}, pp. 51--55, arXiv:nlin.AO/0604026.

\bibitem{Standish05a}
Standish, R.~K. (2005) Complexity of networks. Abbass et~al. (eds.), {\em
  Recent Advances in Artificial Life\/}, Singapore, vol.~3 of {\em Advances in
  Natural Computation\/}, pp. 253--263, World Scientific, arXiv:cs.IT/0508075.

\bibitem{Standish10b}
Standish, R.~K. (2010) Network complexity of foodwebs. Fellerman et~al.
  \cite{ALifeXII}, pp. 337--343.

\bibitem{Standish09a}
Standish, R.~K. (2010) {SuperNOVA}: a novel algorithm for graph automorphism
  calculations. {\em Journal of Algorithms - Algorithms in Cognition,
  Informatics and Logic\/}, submitted, arXix: 0905.3927.

\bibitem{Ulanowicz-etal98}
Ulanowicz, R., Bondavalli, C., and Egnotovich, M. (1998) Network analysis of
  trophic dynamics in {South Florida} ecosystem, {FY} 97: The {Florida Bay}
  ecosystem. Tech. rep., Chesapeake Biological Laboratory, Solomons, MD
  20688-0038 USA, [UMCES]CBL 98-123.

\bibitem{Ulanowicz-etal00}
Ulanowicz, R., Heymans, J., and Egnotovich., M. (2000) Network analysis of
  trophic dynamics in {South Florida} ecosystems, {FY} 99: The graminoid
  ecosystem. Tech. rep., Chesapeake Biological Laboratory, Solomons, MD
  20688-0038 USA, [UMCES] CBL 00-0176.

\bibitem{Watts-Strogatz98}
Watts, D.~J. and Strogatz, S.~H. (1998) Collective dynamics of `small-world'
  networks. {\em Nature\/}, {\bf 393}, 409--10.

\bibitem{Wilhelm-Hollunder07}
Wilhelm, T. and Hollunder, J. (2007) Information theoretic description of
  networks. {\em Physica A\/}, {\bf 385}, 385--396.

\end{thebibliography}

\newpage

\begin{table}
\begin{tabular}{lrrrrrr}
Dataset & nodes & links &${\cal C}$ & $e^{\langle\ln{\cal C}_\mathrm{ER}\rangle}$ & ${\cal
  C}-e^{\langle\ln{\cal C}_\mathrm{ER}\rangle}$ & $\frac{|\ln{\cal
  C}-\langle\ln{\cal C}_\mathrm{ER}\rangle|}{\sigma_\mathrm{ER}}$ \\\hline
celegansneural & 297 & 2345 & 442.7 &251.6 &191.1 &29\\
lesmis & 77 & 508 &199.7 &114.2 &85.4 &24\\
adjnoun & 112 & 850 & 3891 & 3890 & 0.98 & $\infty$\\
yeast & 2112 & 4406 & 33500.6 & 30218.2 & 3282.4 & 113.0\\
celegansmetabolic & 453 & 4050 & 25421.8 & 25387.2 & 34.6&$\infty$  \\
baydry & 128 & 2138 &126.6 &54.2 &72.3 &22\\
baywet & 128 & 2107 &128.3 &51.0 &77.3 &20\\
cypdry & 71 & 641 & 85.7 &44.1 &41.5 &13\\
cypwet & 71 & 632 & 87.4 &42.3 &45.0 &14\\
gramdry & 69 & 911 & 47.4 &31.6 &15.8 &10\\
gramwet & 69 & 912 &54.5 &32.7 &21.8 &12\\
Chesapeake& 39& 177& 66.8& 45.7& 21.1& 10.4\\
ChesLower& 37& 178& 82.1& 62.5& 19.6& 10.6\\
ChesMiddle& 37& 208& 65.2& 48.0& 17.3& 9.3\\
ChesUpper& 37& 215& 81.8& 60.7& 21.1& 10.2\\
CrystalC& 24& 126& 31.1& 24.2& 6.9& 6.4\\
CrystalD& 24& 100& 31.3& 24.2& 7.0& 6.2\\
Everglades &69& 912& 54.5& 32.7& 21.8& 11.8\\
Florida& 128& 2107& 128.4& 51.0& 77.3& 20.1\\
Maspalomas& 24& 83& 70.3& 61.7& 8.6& 5.3\\
Michigan& 39& 219& 47.6& 33.7& 14.0& 9.5\\
Mondego& 46& 393& 45.2& 32.2& 13.0& 10.0\\
Narragan& 35& 219& 58.2& 39.6& 18.6& 11.0\\
Rhode& 19& 54& 36.3& 30.3& 6.0& 5.3\\
StMarks& 54& 354& 110.8& 73.6& 37.2& 16.0\\
PA1 & 100& 99 & 98.9& 85.4 & 13.5& 2.5\\
PA3 & 100& 177 & 225.9 & 207.3& 18.6& 3.0
\end{tabular}
\caption{
  Complexity values of several freely available network
  datasets. celegansneural, lesmis and adjnoun are available from Mark
  Newman's website, 
  representing the neural network of the {\em C. elegans}
  nematode\cite{Watts-Strogatz98}, the coappearance of characters in
  the novel {\em Les Mis\'erables} by Victor Hugo\cite{Knuth93} and
  the adjacency network of common adjectives and nouns in the novel {\em David
  Copperfield} by Charles Dickens\cite{Newman06}. The metabolic data of
  {\em C. elegans}\cite{Duch-Arenas05} and protein interaction network
  in yeast\cite{Jeong-etal01} are available from Duncan Watt's
  website. PA1 and PA3 are networks generated via preferential
  attachment with in degree of one or three respectively, and uniformly
  distributed link weights. The other datasets are food webs 
  available from the Pajek website
  \cite{Christian-Luczkovich99,Ulanowicz-etal98,Ulanowicz-etal00,Hagy02,Baird-etal98,Almunia-etal99,Baird-Ulanowicz89}. For
  each network, the number of 
  nodes and links are given, along with the computed complexity ${\cal
  C}$. In the fourth column, the original network is shuffled 1000
  times, and the logarithm of the complexity is averaged
  ($\langle\ln{\cal C}_\mathrm{ER}\rangle$). The fifth
  column gives the difference between these two values, which
  represents the information content of the specific arrangement of
  links. The final column gives a measure of the significance of this
  difference in terms of the number of standard deviations (``sigmas'') of the
  distribution of shuffled networks. In two examples, the
  distributions of shuffled networks had zero standard deviation, so
  $\infty$ appears in this column.
}
\label{Tab1}
\end{table}

\end{document}